\documentclass[runningheads]{llncs}
\usepackage{graphicx} 
\usepackage{float}
\usepackage{dblfloatfix} 

\usepackage[utf8]{inputenc} 
\usepackage[T1]{fontenc}    
\usepackage{url}            
\usepackage{booktabs}       
\usepackage{nicefrac}       
\usepackage{microtype}      
\usepackage{xcolor}         
\usepackage{comment}

\usepackage{amsmath, amssymb, amscd, bm}
\usepackage{multirow}
\usepackage{nicefrac}       
\usepackage{microtype}      
\usepackage{adjustbox}
\usepackage[ruled,linesnumbered]{algorithm2e}
\usepackage{makecell}
\usepackage{xspace}
\usepackage{enumitem}
\usepackage{caption}

\newcommand{\heading}[1]{\noindent\textbf{#1\xspace}}

\newcommand{\rqq}[1]{
\begin{center}
\begin{tcolorbox}[width=\columnwidth, colback=white!15,left=1pt,right=1pt,top=1pt,bottom=1pt,arc=5pt,auto outer arc]
\textit{#1}
\end{tcolorbox}
\end{center}
}

\newcommand\scalemath[2]{\scalebox{#1}{\mbox{\ensuremath{\displaystyle #2}}}}

\newcommand{\D}{\mathcal{D}}

\newcommand{\X}{\mathcal{X}}
\newcommand{\Y}{\mathcal{Y}}

\newcommand{\btheta}{\boldsymbol\theta}

\newcommand{\bomega}{\boldsymbol\omega}

\DeclareMathOperator*{\argmax}{arg\,max}

\newcommand{\dtoprule}{\specialrule{1pt}{0pt}{\belowrulesep}
            }
\newcommand{\dbottomrule}{
            \specialrule{1pt}{0pt}{\belowrulesep}%
            }

\newcommand{\bx}{\textbf{x}}
\newcommand{\bz}{\textbf{z}}

\newcommand{\bdelta}{\boldsymbol{\delta}}

\def\SP{SP'20\xspace}

\def\ie{\textit{i.e.}\xspace}

\def\eg{\textit{e.g.}\xspace}

\usepackage{tcolorbox}

\usepackage{tikz}
\usepackage{xcolor}

\begin{document}

\title{Bayesian Learned Models Can Detect Adversarial Malware For Free}
 
\author{Bao Gia Doan\inst{1} \and
Dang Quang Nguyen\inst{1}\and
Paul Montague\inst{4} \and 
Tamas Abraham\inst{4} \and 
Olivier De Vel\inst{3} \and
Seyit Camtepe\inst{3} \and 
Salil S. Kanhere\inst{2} \and
Ehsan Abbasnejad\inst{1} \and 
Damith C. Ranasinghe\inst{1}}

\authorrunning{Doan et al.}

\institute{The University of Adelaide, Australia \\
\email{\{giabao.doan, dangquang.nguyen, ehsan.abbasnejad, damith.ranasinghe\}@adelaide.edu.au} \and
The University of New South Wales, Australia \\
\email{salil.kanhere@unsw.edu.au} \and
Data61, CSIRO, Australia \\
\email{seyit.camtepe@data61.csiro.au}, \email{olivierdevel@yahoo.com.au}
\and 
Defence Science and Technology Group, Australia\\
\email{\{paul.montague, tamas.abraham\}@defence.gov.au}}

\maketitle              

\begin{abstract}

Vulnerability of machine learning-based malware detectors to adversarial attacks has prompted the need for robust solutions. Adversarial training is an effective method but is computationally expensive to scale up to large datasets and comes at the cost of sacrificing model performance for robustness. We hypothesize that adversarial malware exploits the low-confidence regions of models and can be identified using epistemic \textit{uncertainty} of ML approaches---epistemic uncertainty in a machine learning-based malware detector is a result of a lack of similar training samples in regions of the problem space. In particular, a Bayesian formulation can capture the model parameters' distribution and quantify epistemic uncertainty without sacrificing model performance. To verify our hypothesis, we consider Bayesian learning approaches with a mutual information-based formulation to quantify uncertainty and detect adversarial malware in \textit{Android}, \textit{Windows} domains and \textit{PDF} malware. We found, quantifying uncertainty through Bayesian learning methods can defend against adversarial malware. In particular, Bayesian models: (1)~are generally capable of identifying adversarial malware in both feature and problem space, (2)~can detect concept
drift by measuring uncertainty, and (3)~with a diversity-promoting approach (or \textit{better posterior approximations}) leads to parameter instances from the posterior to significantly enhance a detectors' ability. 
\end{abstract}

\keywords{Malware Detection  \and Adversarial Malware \and Bayesian Learning}

\section{Introduction}

The world is witnessing an alarming surge in malware incidents causing significant damage on multiple fronts. Financial costs are reaching billions of dollars~\cite{anderson2019measuring} and as highlighted in~\cite{eddy_perlroth_2020}, human lives are also at risk. At the end of 2023, Kaspersky Lab reported that an average of 411,000 malware instances were detected each day~\cite{kaspersky}. Addressing widespread malware attacks is an ongoing challenge, and prioritizing research to develop automated and efficient systems for detecting and combating malware effectively is essential.

Recent advances in Machine Learning (ML) have led to highly effective malware detection systems~\cite{arp2014drebin,peng2012using,harang2020sorel,anderson2018ember,raff2018malware}. However, ML-based models are susceptible to attacks from \textit{adversarial examples}. Initially observed in the field of computer vision~\cite{goodfellowExplainingHarnessingAdversarial2015,pgd,biggio2018wild}, this vulnerability extends to the domain of malware detection, giving rise to so-called \textit{adversarial malware}~\cite{pierazzi2020intriguing,grosse2017adversarial,kolosnjaji2018adversarial,kreuk2018deceiving,doan2023featurespace}. These attacks involve carefully modifying malware samples to \textit{retain their functionality} and {realism} while making minimal changes to the underlying code. Consequently, attackers can deceive ML-based malware detectors by misguiding them to misclassify the adversarial malware as benignware. The emergence of such attacks poses a significant and evolving threat to ML-based malware detection systems, as highlighted in recent studies ~\cite{suciu2019exploring,pierazzi2020intriguing,demetrio2021secml,doan2023featurespace}.

\vspace{2mm}
\noindent\textbf{Problem.~}
In general, to defend against adversarial examples, adversarial training~\cite{athalye2018obfuscated} is an effective method.  But:
\begin{itemize}
    \item Generating adversarial malware samples for training, especially with large-scale datasets (typical in the malware domain) for deployable models, is shown to be non-trivial~\cite{pierazzi2020intriguing,doan2023featurespace}). Fast, gradient-based methods to craft perturbations to construct adversarial malware in the discrete space of software code binaries (\textit{problem space}) from vectorized features (\textit{feature space}) is difficult. Because the function mapping from the problem space to the features is non-differentiable~\cite{biggio2013evasion,biggio2013security}.
    \item It is difficult to enforce and maintain functionality, realism and maliciousness constraints in a scalable and automated manner to generate adversarial malware in the problem space. 
    For instance, the transformations used in~\cite{yang2017malware} led to app crashes as most malware could not function after manipulation.
\end{itemize}

\begin{figure}[b!]
    \centering
    \includegraphics[width=.9\linewidth]{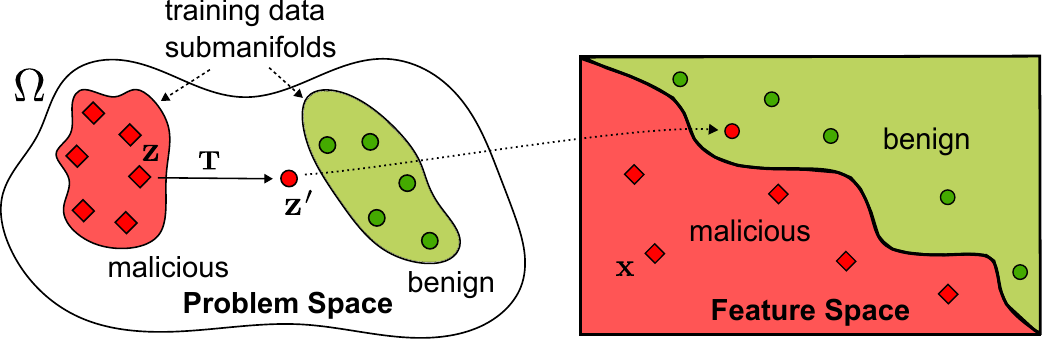}
    \caption{Illustration of \textit{functional}, \textit{realistic}, adversarial malware in the problem space, where $\bz'$ is the transformation of $\bz$ (a malware app) that passes the decision boundary in the detector's feature space and successfully fools the malware detector whilst satisfying problem-space constraints $\Omega$. The white areas, outside of the training data submanifolds, are regions of high uncertainty for ML-based malware detectors. }
    \label{fig:manifold}
    \vspace{-5mm}
\end{figure}

Interestingly, a recent study shows the projection of perturbed yet functional malware in the problem space (the discrete space of software code binaries) into the feature space will be a subset of feature-space adversarial examples~\cite{doan2023featurespace}. So, an adversarially trained network with feature-space adversarial samples is inherently robust against problem-space adversarial malware. But:

\rqq{A significant problem with adversarial training, besides the problem of generating adversarial malware and the increased cost of training a network, is the compromise in model performance necessary to achieve robustness. 
The challenge of achieving robustness without compromising detector performance presents an intricate trade-off.
}

\noindent\textbf{Research Questions.~}
In contrast to adversarial learning for robustness, we investigate a different approach. As illustrated in Figure~\ref{fig:manifold}, given training is always data limited to some submanifolds, we argue that adversarial malware exploits the low-confidence regions of ML-based models as attackers seek the minimal transformation (\textbf{T}) needed to move a model decision from malware to benignware. Because, adversarial malware construction is constrained by functional requirements; arbitrary changes to binaries are not possible and will break the malware code. Consequently:

\rqq{We hypothesize adversarial malware could be detected by analyzing the epistemic uncertainty captured and expressed by ML malware detectors.}

Epistemic uncertainty in machine learning-based malware detectors results from a lack of similar training samples in regions of the problem space. We argue, it is these problem space regions that an adversary seeks to exploit in their pursuit of functional, realistic adversarial malware. Exploiting uncertainty itself is not new, but our contributions arise from investigating \textit{practical} methods for, \textit{both}, capturing and expressing epistemic uncertainty and evaluating their \textit{efficacy} in the detection of \textit{problem space malware}. The efficacy of such an uncertainty-based defense against \textit{adversarial malware}---adversarial examples in the malware domain---remains to be understood. So, in this study, we seek to validate our hypothesis by answering the following research questions (\textit{RQs}):

\begin{itemize}[]
    \item[] \textit{\textbf{RQ1}:} How can we \textit{practically} capture epistemic uncertainty in malware detection tasks?
    \item[] \textit{\textbf{RQ2}:} How \textit{effective} are uncertainty measures, in general, in detecting adversarial malware?
    \item[] \textit{\textbf{RQ3}:} How well does quantifying uncertainty to detect adversarial malware \textit{generalize} across malware domains?
\end{itemize}

\noindent\textbf{Our Approach.~}To address the questions we posed, we investigate practical approaches to capture and measure uncertainty in ML-based malware detection tasks. 
We realize formulations in the context of Bayesian deep neural networks preserve uncertainty. Specifically, Bayesian deep learning methods infer the distribution of model parameters to realize robust models and express epistemic uncertainty through the predictions sampled from each parameter particle to a given input. Unfortunately, the exact inference of a parameter distribution in the context of deep learning is intractable. Therefore, we propose exploring the approximation of Bayesian neural networks (BNNs)~\cite{Liu2016,li2017dropout,blundell2015weight} able to scale up to large and complex malware datasets to measure uncertainties. Whilst Bayesian models can directly express predictive uncertainty as well as model predictions, we explore the formulation of mutual information for quantifying epistemic uncertainty possible in the context of Bayesian models. 

We found epistemic uncertainty: i)~captured by Bayesian deep neural networks able to approximate the posterior better; and ii)~quantified by mutual information, is highly effective in detecting adversarial malware. Further, the approach is: i)~\textit{free} (the epistemic uncertainty inherently exists in BNNs and  \textit{adversarial malware} detection is improved without compromising detection performance); and ii)~ very versatile---\textit{i.e.} adaptable to various deep neural networks in different malware domains, including \textit{Android} and \textit{Windows} Portable Executable (PE) files and \textit{PDF} malware.

\vspace{1mm}
\noindent\textbf{Our Contributions.~}
\begin{enumerate}
    \item We propose a \textit{practical} and \textit{effective} approach to detect adversarial malware without needing to sacrifice model performance. \item To detect adversarial malware, we leverage Bayesian learning to capture epistemic uncertainty and employ a mutual information formulation for expressing uncertainty in the context of Bayesian neural networks.  
    \item Through extensive experiments, we show the proposed method's \textit{generalizability} and \textit{effectiveness} in detecting adversarial malware across both \textit{the problem space} and \textit{feature space} as well as across  malware domains, including Android, Windows and PDF malware.  
\end{enumerate}

Importantly, our findings show Bayesian learned models able to better approximate the posterior (\textit{model distribution}) is highly effective at detecting both problem space and adversarial malware.

\section{Background and Related Work}

\vspace{1mm}

\noindent\textbf{Adversarial Malware.} Research on Android malware detection primarily addresses adversarial attacks, including query-based evasion~\cite{croce2022sparse}, gradient-based evasion~\cite{li2020adversarial,li2021framework}, and feature modification-based evasion~\cite{demontis2017yes,grosse2017adversarial}. These attacks extract slices of bytecodes from benign apps~\cite{pierazzi2020intriguing,yang2017malware}, use obfuscation tools~\cite{demontis2017yes}, or modify dummy codes like unused API calls~\cite{chen2019android}.

Another approach involves problem-space transformations to generate realistic adversarial malware, guided by feature-space perturbations. These transformations adhere to constraints like preserved semantics and plausibility~\cite{pierazzi2020intriguing}. For instance, \cite{pierazzi2020intriguing} proposed an evasion attack creating real-world adversarial Android apps through such transformations. Other techniques include evolution and confusion attacks~\cite{yang2017malware} and obfuscation~\cite{demontis2017yes} for manipulating Android malware.

\vspace{2px}
\noindent\textbf{Measuring Uncertainties.}\cite{feinman2017detecting} explores model confidence on adversarial samples in Computer Vision (CV) by examining Bayesian uncertainty estimates using prediction variance. Similarly, \cite{smith2018understanding} investigates uncertainty measures like Mutual Information (MI) for detecting adversarial examples in the CV domain.

In the malware detection domain, limited research focuses on leveraging uncertainty. \cite{backes17} and \cite{nguyenLeveragingUncertaintyImproved2021} propose leveraging uncertainty in Android malware analysis to reduce incorrect decisions. However, they don't quantify uncertainty for adversarial malware. \cite{li2021can} finds that models preserving uncertainty are useful for detecting dataset drifts but struggle with adversarial examples.

Existing malware research often overlooks the impact of chosen uncertainty quantification measures. While measures like mutual information~\cite{rawat2017adversarial} and predictive entropy~\cite{smith2018understanding} exist, research in adversarial attacks on malware, a domain with unique characteristics, is lacking. The malware domain requires maintaining functionality, and malware evolves rapidly over time, presenting a distinct challenge not addressed by current research.

\vspace{1mm}
\noindent\textbf{Summary.~}We recognize that: i) extensive and quantitative investigations of the capability and practicability of various uncertainty measures from diverse Bayesian learning to detect realistic, functional adversarial malware have not been performed; ii) the effectiveness and generalization of this manner of approach across the malware domain is unclear.

\section{Problem Definition}

\subsection{Threat Model}

In this paper, we focus on \textit{evasion} attacks. The threat model of this attack is described below:
\begin{itemize}
    \item \textbf{Adversary's Goal.} 
    The adversary aims to manipulate the Android malware detector in such a way that it incorrectly classifies the adversarial (malware) example as benign.
    
    \item \textbf{Adversary's Knowledge.} In this study, we focus on an adversary who possesses perfect knowledge (PK)~\cite{biggio2013evasion}. This type of attacker possesses comprehensive knowledge, including all target model parameters, its learning algorithm, training data, and parameters. This knowledge is utilized to create adversarial malware.

    \item \textbf{Adversary's Capability.} The adversary has the capability to craft adversarial malware through two different attack spaces. The first involves manipulating feature representations within specific constraints in the \textit{feature space}~\cite{pgd}. The second entails applying a series of transformations while adhering to \textit{problem-space} domain constraints~\cite{pierazzi2020intriguing}.
    
\end{itemize}

\vspace{-2mm}
\subsection{Adversarial Malware Attacks}

\noindent\textbf{Problem-Space Attacks.~}
The problem space $\mathcal{Z}$ corresponds to the input space of real objects in a specific domain, such as software binaries. To process the problem space using machine learning (ML), it is necessary to transform $\mathcal{Z}$ into a compatible format, typically numerical vector data~\cite{arp2014drebin}. This transformation is achieved through a feature mapping function $\Phi: \mathcal{Z} \rightarrow \mathcal{X} \subseteq \mathbb{R}^n$, which maps a software binary $\mathbf{z} \in \mathcal{Z}$ to an $n$-dimensional feature vector $\mathbf{x} \in \mathcal{X}$ in the feature space ($\Phi(\mathbf{z}) = \mathbf{x}$).
These features are then learned by an ML-based network, generally defined as a function $f: \mathcal{X} \rightarrow \mathcal{Y}$ parametrized by a set of weights and biases denoted by $\btheta$. 

In the context of adversarial malware attacks, attackers typically apply a transformation to the problem space object $\mathbf{z}$, resulting in a modified object $\mathbf{z}'$ that is mapped to a feature vector $\mathbf{x}'$ close to the target feature vector in the feature space. Formally, given a problem-space object $\mathbf{z} \in \mathcal{Z}$ with label $y \in \mathcal{Y}$, the goal of the adversary is to find a transformation function $\mathbf{T}: \mathcal{Z} \rightarrow \mathcal{Z}$ (e.g., addition, removal, modification) such that the transformed object $\mathbf{z}' = \mathbf{T}(\mathbf{z})$ is classified as a different class, \ie $\argmax p(y \mid \Phi(\mathbf{T}(\bz')), \btheta) = t \neq y$, while satisfying the problem-space constraints (available transformations, preserved semantics, plausibility, robustness to pre-processing~\cite{pierazzi2020intriguing}) denoted by $\Omega$ as shown in Figure~\ref{fig:manifold}. 

\vspace{1mm}
\noindent\textbf{Feature-Space Attacks.~}
We note that feature-space attacks are well defined and consolidated in related work~\cite{biggio2018wild,carlini2017towards,grosse2017adversarial}. In this paper, we use a popular feature mapping function provided in the DREBIN~\cite{arp2014drebin} and EMBER~\cite{anderson2018ember} dataset to map raw bytes of software to a vector of $n$ features for Android and Windows malware respectively. 
A feature-space attack is then to modify a feature-space object $\bx \in \X$ to become $\bx' = \bx + \bdelta$ where $\bdelta$ is the added perturbation crafted with an \textit{attack objective function} to misclassify $\bx'$ into another class, \ie $\argmax p(y \mid \bx', \btheta) = t \neq y$ where $y \in \Y$ is the ground-truth label of $\bx$. 
We note that in the malware domain (a binary classification task), the attackers' goal is to make the malware be recognized as benignware. These modifications have to follow feature-space constraints.
We denote the constraints on feature-space modifications by $\Upsilon$.
Given a sample $\bx \in \X$, the feature-space modification, or perturbation $\bdelta$ must satisfy $\Upsilon$. This constraint $\Upsilon$ reflects the realistic requirements of problem-space objects. Malware feature perturbations $\bdelta$ can be constrained as $\bdelta_{lb} \leq \bdelta \leq \bdelta_{ub}$~\cite{pierazzi2020intriguing}.

\section{Measuring Uncertainty}

\vspace{-2mm}
This paper proposes using uncertainty as a measure for detecting adversarial malware. 
The proposed method involves training a model capable of capturing predictive uncertainty. This uncertainty level is then employed as a measure to identify potential adversarial samples, with higher uncertainty indicating a greater likelihood of being adversarial.

It is crucial to highlight a common misunderstanding in classification models. People often mistake the final probability vector obtained from regular deterministic networks (usually after applying the \textit{softmax} function to the last layer of the neural network classifier) as an accurate measure of the model's \textit{confidence}. However, it is essential to recognize that a model can still have significant uncertainty (low confidence) in its predictions, even if it produces a high softmax output (e.g., 100\%)~\cite{gal2016uncertainty}.

On the other hand, confidence naturally arises from uncertainty present in models such as Bayesian models. Hence, this study uses Bayesian neural networks to leverage their inherent uncertainty to detect adversarial malware \textit{for free}.

\subsection{Bayesian Machine Learning for Malware Detection}
\label{sec:bayes}

In general, we assume a set $D$ of $n$ training examples $(\bz_i, y_i)$ with binary outputs. The ML-based detectors first map the inputs $\bz$ to feature-space vectors $\bx=\Phi(\bz)$. These feature-space vectors are then utilized by ML-based techniques such as Deep Neural Networks (DNNs) to discriminate between benignware and malware.

Instead of considering the parameters ($\btheta$) as fixed to be optimized, the Bayesian approach considers them as random variables. Thus, a prior distribution \(p(\btheta)\) is assigned to the weights of the network. By also having a likelihood function \(p(y \mid \mathbf{x}, \btheta)\), which represents the probability of obtaining \(\mathbf{y} \in \mathcal{Y}\) given a specific set of parameter values $\btheta$ and an input to the network $\bx$, it becomes possible to perform inference on a dataset by marginalizing the parameters. 
Thus, the goal of Bayesian learning is to find the posterior distribution using Bayes theorem: 
\begin{equation*}
p(\btheta\mid\D)={\prod_{(\bx,y)\sim\D}p(y\mid\bx,\btheta)p(\btheta)}/{Z}
\end{equation*}
where $Z$ is the normalizer, $\D$ is training dataset.

The complex, high-dimensional, and non-convex nature of the posterior in Bayesian neural networks renders direct estimation infeasible, necessitating the use of approximation techniques. Among these, the Laplace approximation~\cite{mackay1992practical,ritter2018scalable}, Dropout~\cite{li2017dropout,srivastava2014dropout}, Variational Inference~\cite{blundell2015weight}, and Stein Variational Gradient Descent (SVGD)~\cite{Liu2016} stand out as practical approximation methods. Although SVGD, particularly with repulsive force, shows promise for better posterior approximation~\cite{d2021stein,doan2022bayesian}, we also investigate Dropout and Variational Inference along with general ensembles as viable and different approximation alternatives.

\vspace{2px}
\heading{Variational Inference (VI).~}
The concept of Variational Inference (VI) involves approximating the intractable posterior \(p(\btheta \mid \mathcal{D})\) with a simpler approximate distribution \(q_{\bomega}(\btheta)\). The objective is to maximize the evidence lower bound (ELBO) as follows:
\[
\mathcal{L}_{VI} := \int q_{\bomega}(\btheta) \log p(\mathcal{D} \mid \btheta) d\btheta - D_{KL}(q_{\bomega} \mid\mid p(\btheta)).
\]
The advantage of this method lies in transforming the typically intractable Bayesian inference problem into an optimization challenge of maximizing a parameterized function, amenable to standard gradient-based techniques. The variational inference (VI) technique simplifies the process by replacing fixed weights with parameters like means and standard deviations (assuming a Gaussian distribution).

\heading{Dropout.~}Another widely used method for approximating Bayesian neural networks is Dropout~\cite{srivastava2014dropout}. Dropout involves randomly setting the outputs of neural network units to zero, effectively creating multiple variations of the network. This generates an approximation of the posterior distribution using a Monte Carlo (MC) estimator~\cite{li2017dropout}: 
\begin{equation*}
\begin{aligned}
\mathbb{E}_{p(\btheta | \mathcal{D})}[f^{\btheta}(\bx)] &= \int p(\btheta | \mathcal{D}) f_{\btheta}(\bx) \text{d} \btheta \simeq \int q_{\bomega}(\btheta) f_{\btheta}(\bx) \text{d} \btheta \simeq \frac{1}{n} \sum_{i=1}^{n} f_{\btheta_i}(\bx), \; \btheta_{1..n} \sim q_{\bomega}(\btheta).
\end{aligned}
\end{equation*}

Using this Dropout technique~\cite{li2017dropout,d2021stein}, we only need to add Dropout layers into the neural networks, and we can approximate the posterior distribution during the inference/validation phase by randomly dropping out neurons and using the Monte Carlo estimator mentioned above. 

\vspace{2px}
\heading{Stein Variational Gradient Descent (SVGD).~} An alternative method for posterior approximation is SVGD~\cite{Liu2016}. This method has several advantages. Firstly, it learns multiple \emph{network parameter particles} in parallel, which leads to faster convergence. Secondly, it has a \emph{repulsive factor} that encourages the diversity of parameter particles, helping to prevent mode collapse - a challenge in posterior approximation. Thirdly, unlike the aforementioned methods, it does not need any modification to neural networks, making it easy to adapt to existing neural networks. 

This approach considers $n$ samples from the posterior (\ie parameter particles). The variational bound is minimized when gradient descent is modified as:
\begin{equation}
\label{eq:grad_update}
\btheta_i = \btheta_i - \frac{\epsilon_i}{n} \sum_{j=1}^n \left[ k(\btheta_j, \btheta) \nabla_{\btheta_j} \ell(f_{\btheta_j}(\bx),y) - \gamma\nabla_{\btheta_j} k(\btheta_j, \btheta) \right]
\end{equation}

 Here, $\btheta_i$ is the $i$th particle, $n$ is the number of particles, $k(\cdot, \cdot)$ is a kernel function that measures the similarity between particles, and $\gamma$ is a hyper-parameter.
 Thanks to the kernel function, the parameter particles are encouraged to be dissimilar to capture more diverse samples from the posterior. This is controlled by a hyper-parameter $\gamma$ to manage the trade-off between diversity and loss minimization. Following~\cite{Liu2016}, we use the RBF kernel $\scalemath{.9}{k(\btheta, \btheta')=\exp \left(-{\left\|\btheta-\btheta^{\prime}\right\|^{2}}/{2 h^{2}}\right)}$ and take the bandwidth $h$ to be the median of the pairwise distances of the set of parameter particles at each training iteration.

\vspace{2mm}
\noindent\textbf{Prediction.~}Regardless of the above-mentioned Bayesian approaches, at the prediction stage, given the test data point $\bx^*$, we can obtain the prediction by approximating the posterior using the Monte Carlo samples as:
{\small\begin{align}
    p ( y^* \mid & \mathbf{x}^*, \D ) = \int p(y^* \mid \mathbf{x}^*, \boldsymbol{\theta}) p(\boldsymbol{\theta} \mid \D) d \boldsymbol{\theta} 
    & \approx \frac{1}{n} \sum_{i=1}^{n} p(y^* \mid \mathbf{x}, \boldsymbol{\theta}_{i}), \quad \boldsymbol{\theta}_{i} \sim p(\boldsymbol{\theta} \mid \D)\,
    \label{eq:mc}
\end{align}}
where $\btheta_i$ is an individual parameter particle. 
Note that we hypothesize that it is critical to have diverse parameter particles, as this will promote uncertainty when dealing with adversarial malware.

\subsection{Uncertainty Measures}

Given the above-mentioned Bayesian approximations, we can now leverage the Bayesian approach to attain uncertainty measures from Bayesian models: 

\vspace{2mm}
\heading{Predictive Entropy (PE).~}In the malware classification tasks, where the output of a malware detector is a conditional probability distribution $P(y\mid\bx)$ over some discrete set of outcomes $\mathcal{Y}$, we can obtain the uncertainty by leveraging the entropy of the predictive distribution, \ie \textit{predictive entropy}: 
\begin{equation}
    \label{eq:pe}
    H[p(y \mid \D, \bx)]=-\sum_{y \in \mathcal{Y}} p(y \mid \D,\bx) \log p(y \mid \D,\bx)
    \vspace{-3mm}
\end{equation}
One advantage of this measure is that it can be applied even on deterministic neural networks. For Bayesian networks, $p$ is approximated using the MC approach, as in Equation~\ref{eq:p_mc}.

\vspace{2mm}
\heading{Mutual Information (MI):} MI quantifies the information gain about the model's parameters, denoted as $\boldsymbol{\theta}$, upon observing new data. It measures the reduction in uncertainty about $\boldsymbol{\theta}$ when a label $y$ is obtained for a new malware sample $\mathbf{x}$, given the pre-existing dataset $\mathcal{D}$. MI between the model parameters and the new data can be mathematically represented as follows:
\begin{align*}
  MI(\boldsymbol{\theta}; y | \mathcal{D}, \mathbf{x}) &= H[y | \mathcal{D}, \mathbf{x}] - \mathbb{E}_{p(\boldsymbol{\theta} | \mathcal{D})}[H[y | \boldsymbol{\theta}, \mathbf{x}]],
\end{align*}
where $H[y | \mathcal{D}, \mathbf{x}]$ denotes the entropy of the predictive distribution over the label $y$ given the new sample $\mathbf{x}$ and the dataset $\mathcal{D}$. The term $\mathbb{E}_{p(\boldsymbol{\theta} | \mathcal{D})}[H[y | \boldsymbol{\theta}, \mathbf{x}]]$ represents the expected value of the conditional entropy of $y$ given the model parameters $\boldsymbol{\theta}$ and the new sample $\mathbf{x}$, averaged over the posterior distribution of $\boldsymbol{\theta}$ given $\mathcal{D}$.
From the above definition, MI essentially measures the model's \textit{epistemic} uncertainty. If the parameters at a point are well defined (\eg data seen during training), then we would gain little information from the obtaining label, or the MI is low. 
This characteristic is crucial since it can aid in detecting adversarial malware; however, it is currently absent in most literature.

Notably, all of these quantities are usually intractable in deep neural networks; however, we can approximate them using Monte Carlo. In particular, 
\begin{align}
  p(y \mid \mathcal{D}, \mathbf{x}) &\simeq \frac{1}{n} \sum_{i=1}^{n} p(y \mid \btheta_i, \mathbf{x}) \label{eq:p_mc}
                                    := p_{MC}(y \mid \D, \mathbf{x}) \nonumber \\
  H[ p(y \mid \mathcal{D}, \mathbf{x}) ] &\simeq H[ p_{MC}(y \mid \mathcal{D}, \mathbf{x})] \\
  I(\btheta, y \mid \mathcal{D}, x) &\simeq H[ p_{MC}(y \mid \mathcal{D}, \mathbf{x})] - \frac{1}{n} \sum_{i=1}^{n}  H[p(y \mid \btheta_i, \mathbf{x})]
\end{align}

In the following section, we will empirically study these above-mentioned uncertainty measures.

\section{Experiments and Results}

\subsection{Experimental Setup}
\label{sec:exp-setup}

We implement the experiments using PyTorch~\cite{paszke2019pytorch}, SecML~\cite{melis2019secml} and Bayesian-Torch~\cite{krishnan2022bayesiantorch} libraries and run experiments on a CUDA-enabled GTX A6000 GPU. Below are details of the datasets and classifiers. 

\vspace{1mm}
\noindent\textbf{Malware Classifiers.~}
We utilize the Feed-Forward Neural Network (FFNN) provided in~\cite{harang2020sorel}. This network architecture is utilized in Android and Windows malware, as well as in PDF malware detection tasks. 
Our network implementation uses the default configuration provided in~\cite{harang2020sorel}. We also adopt the architecture of FFNN to design the Bayesian Neural Network (BNN).

\vspace{2mm}
\heading{Inference.~}
Below is the detailed implementation for each of the inference approaches. We utilize a number of inference $n=10$ following previous research~\cite{li2021can} across all methods for a fair comparison:
\begin{itemize}
    \item \textit{MC Dropout.~} We add dropout layers into fully-connected layers of neural networks with a dropout rate of 0.5. In the inference phase, the network is forward-passed 10 times for each sample to estimate the posterior. 
    \item \textit{VI.~}We sample 10 parameters of the fully-connected layer (\ie weights and biases) from Gaussian distributions. The mean and standard deviation variables of Gaussian distributions are learned via back propagation using the reparameterization technique~\cite{blundell2015weight}, and we use the implementation from Bayesian Torch~\cite{krishnan2022bayesiantorch}.
    \item \textit{SVGD.~}We train 10 different parameter particles in parallel using the objective mentioned in Section~\ref{sec:bayes}. We also sample 10 predictions for each malware sample in the inference phase for consistency with other Bayesian approaches.
    \item \textit{Ensemble.~}We trained 10 malware detectors with random seeds and used them in an ensemble prediction to compare with Bayesian approaches.
\end{itemize}

\noindent\textbf{Dataset.}
We use a public Android dataset~\cite{pierazzi2020intriguing} based on the DREBIN feature space~\cite{arp2014drebin}, a binary feature set widely employed in recent research~\cite{pierazzi2020intriguing,li2021framework}. The dataset, spanning January 2017 to December 2018, includes approximately 152K Android apps with $\sim$ 135K benign and $\sim$ 15K malicious apps. An app is labeled malicious if detected by four or more VirusTotal AVs. For Windows, we use the popular EMBER~\cite{anderson2018ember} dataset, including pre-extracted samples of Windows apps. In addition, we also employ the Contagio dataset~\cite{Parkour} for PDF malware with $\sim$ 17K clean and $\sim$ 12K malicious PDFs.

\vspace{2px}

\noindent\textbf{Attacks.}
In this paper, we concentrate on realistic attacks, where evasion attacks adhere to \textit{problem-space constraints} for realism and functionality. We utilize the \SP attack from~\cite{pierazzi2020intriguing}, a white-box attack producing realistic adversarial malware within these constraints. Due to high computational complexity, we generate a set of problem-space adversarial malware from the \SP attack using the released codebase and evaluate our approach's robustness. Additionally, we consider \textit{feature-space} adversarial attacks, a superset of realistic adversarial malware according to recent research~\cite{doan2023featurespace}. We employ the PGD L1 attack~\cite{pgd} as well as BCA~\cite{al2018adversarial} and Grosse~\cite{grosse2016adversarial} feature-space attacks to demonstrate the effectiveness of our proposed method on DREBIN features \cite{arp2014drebin}. 

For Windows malware, regarding \textit{problem-space constraints}, we use the adversarial malware set released by~\cite{erdemir2021adversarial}. This set leverages the method of~\cite{Fleshman}, the winner of the machine learning static evasion competition~\cite{DEFCON}. Moreover, we also leverage the \textit{feature-space} attacks, namely, the unbounded gradient attack \cite{pdfclassifier} method for the PDF malware.

\vspace{2px}
\noindent\textbf{Evaluation Metrics.} We present the performance of classifiers in detecting malware under two scenarios: i) clean performance without attacks and ii) resilience against evasion attacks with adversarial malware. Metrics include AUC, F1, Precision, and Recall. The ROC curve evaluates on a set with benignware from the test set (negative examples) and adversarial malware generated using attacks like the problem-space \SP attack~\cite{pierazzi2020intriguing} or feature-space PGD attack~\cite{pgd} as positive samples.

\subsection{Clean Performance (No Attacks) in Android Domain}

First, we aim to evaluate the performance of networks in an Android malware detection task. The results in Table~\ref{tab:clean-performance-android} indicate that all the networks under consideration are proficient in detecting malware, with an AUC exceeding 90\%.

\vspace{-3mm}
\begin{table}[h]
\centering
\setlength{\tabcolsep}{8pt}
\resizebox{.55\linewidth}{!}{%
\begin{tabular}{@{}lllll@{}}
\toprule
Networks         & F1    & Precision & Recall  & AUC\\ \midrule
FFNN    &  94.52\% & 97.21\%     & 93.12\%  & 96.42\%\\ 
MC Dropout &  93.52\% & 94.97\%     & 92.45\%  & 95.72\%\\
ELBO &  93.37\%  & 95.47\%     & 91.76\%  & 95.18\%\\ 
Ensemble &  94.82\%  & 97.56\%     & 93.52\%  & 96.89\%\\ 
SVGD &  93.45\%  & 96.23\%     & 91.68\%  & 95.48\%\\ 
\bottomrule
\end{tabular}%
}
\vspace{2mm}
\caption{The clean performance of various models in Android malware detection task (FFNN is \textit{non-Bayesian} baseline).}
\label{tab:clean-performance-android}
\end{table}
\vspace{-11mm}

\subsection{Robustness against Problem-Space Adversarial Android Malware}
\vspace{-1mm}

\begin{figure}[h]
    \centering
    \includegraphics[width=1.0\linewidth]{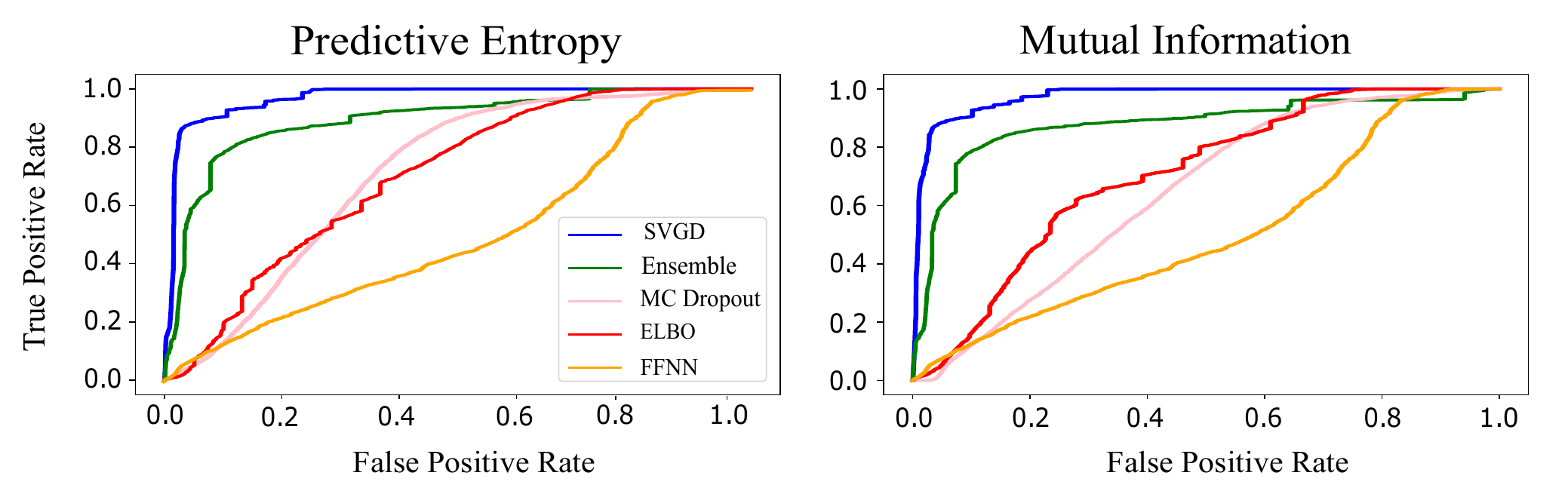}
     \caption{Using mutual information and predictive entropy to detect \textit{problem-space} Android adversarial malware from \SP attacks with a budget $\epsilon=90$ (FFNN is a \textit{non-Bayesian} baseline).}
    \label{fig:android-mi-pe}
    \vspace{-3mm}
\end{figure}

In this section, we evaluate the robustness of our approach against one of the state-of-the-art \textit{problem-space} attacks in the field, conducted in the \SP paper \cite{pierazzi2020intriguing}. To conduct the \SP attack \cite{pierazzi2020intriguing}, we crawl real APK files from Androzoo corresponding to the True Positive Samples of the base network. In total, we gathered more than 4.6K of real Android malware to generate adversarial samples. We evaluate the effectiveness of evaluated networks against attacks with increasing attack budgets ($\epsilon$ from 30 to 90). Table 3 shows that Bayesian versions generally perform better than a single FFNN, while the diversity-promoting Bayesian approach (SVGD) outperforms the rest, with AUC higher than 96\% across all tested attacking budgets. 
A visualized AUC curve for the attack budget of $\epsilon$ = 90 is shown in Figures \ref{fig:android-mi-pe} for Mutual Information and Predictive Entropy, respectively.

\begin{table}[h]
\centering
\setlength{\tabcolsep}{5pt}
\resizebox{.95\linewidth}{!}{%
\begin{tabular}{cccccccccccc}
\dtoprule
\multirow{2}{*}{\textbf{\makecell{Networks/\\Attacks}}} & \multirow{2}{*}{$\epsilon$} & \multicolumn{2}{c}{FFNN} & \multicolumn{2}{c}{Dropout} & \multicolumn{2}{c}{ELBO} & \multicolumn{2}{c}{Ensemble} & \multicolumn{2}{c}{SVGD}            \\ 
                                          &                      & PE            & MI       & PE           & MI           & PE          & MI         & PE            & MI           & PE               & MI               \\ \midrule
\multirow{3}{*}{\SP}                      & 30                    & 69.62\%       & NA       & 75.61\%      & 67.9\%      & 63.18\%     & 76.61\%    & 93.63\%       & 93.82\%      & \textbf{98.02\%} & \textbf{98.48\%} \\ 
                                          & 60                   & 50.16\%       & NA       & 71.34\%      & 64.32\%      & 75.03\%     & 73.61\%    & 89.31\%       & 87.71\%      & \textbf{96.91\%} & \textbf{97.33\%} \\ 
                                          & 90                   & 52.72\%       & NA       & 70.53\%      & 63.12\%      & 77.39\%     & 74.65\%    & 87.54\%       & 89.15\%      & \textbf{96.82\%} & \textbf{97.27\%} \\ \dbottomrule
\end{tabular}%
}
\vspace{2mm}
\caption{Detection performance against \textit{problem-space} adversarial malware from \SP attacks (FFNN is a \textit{non-Bayesian} baseline).}
\label{tab:sp-results}
\vspace{-5mm}
\end{table}

\subsection{Robustness against Feature-Space Adversarial Android Malware}

\vspace{2px}
Problem-space attacks are known to be a subset of feature-space attacks~\cite{doan2023featurespace}. Thus, in this section, we want to validate the method's effectiveness against feature-space attacks. In particular, we use Projected Gradient Descent (PGD) attacks, one of the prevalent feature-space attacks. For a fair comparison, both problem-space and feature-space attacks are bounded by the same L1 norm, $\epsilon$.

\begin{table}[h]
\centering
\setlength{\tabcolsep}{5pt}
\resizebox{.95\linewidth}{!}{%
\begin{tabular}{cccccccccccc}
\dtoprule
\multirow{2}{*}{\textbf{\makecell{Networks/\\Attacks}}} & \multirow{2}{*}{$\epsilon$} & \multicolumn{2}{c}{FFNN} & \multicolumn{2}{c}{Dropout} & \multicolumn{2}{c}{ELBO} & \multicolumn{2}{c}{Ensemble} & \multicolumn{2}{c}{SVGD}            \\ 
                                          &                      & PE            & MI       & PE           & MI           & PE          & MI         & PE            & MI           & PE               & MI               \\ \midrule
\multirow{3}{*}{PGD-L1}                      & 30                    & 13.56\%       & NA       & 14.86\%      & 17.74\%      & 15.65\%     & 18.56\%    & 72.21\%       & 74.54\%      & \textbf{97.01\%} & \textbf{97.62\%} \\ 
                                          & 60                   & 12.34\%       & NA       & 13.45\%      & 14.81\%      & 14.21\%     & 16.95\%    & 65.32\%       & 66.01\%      & \textbf{97.15\%} & \textbf{97.73\%} \\ 
                                          & 90                   & 12.23\%       & NA       & 14.35\%      & 15.85\%      & 14.73\%     & 16.75\%    & 51.12\%       & 54.34\%      & \textbf{97.32\%} & \textbf{97.85\%} \\ \dbottomrule
\end{tabular}%
}
\vspace{1mm}
 \caption{Detection performance against PGD-L1 \textit{feature-space} adversarial malware(FFNN: \textit{non-Bayesian} baseline).}
\label{tab:pgd-results}
\end{table}

\vspace{-8mm}
\begin{figure}[h]
    \centering
    \includegraphics[width=1.0\linewidth]{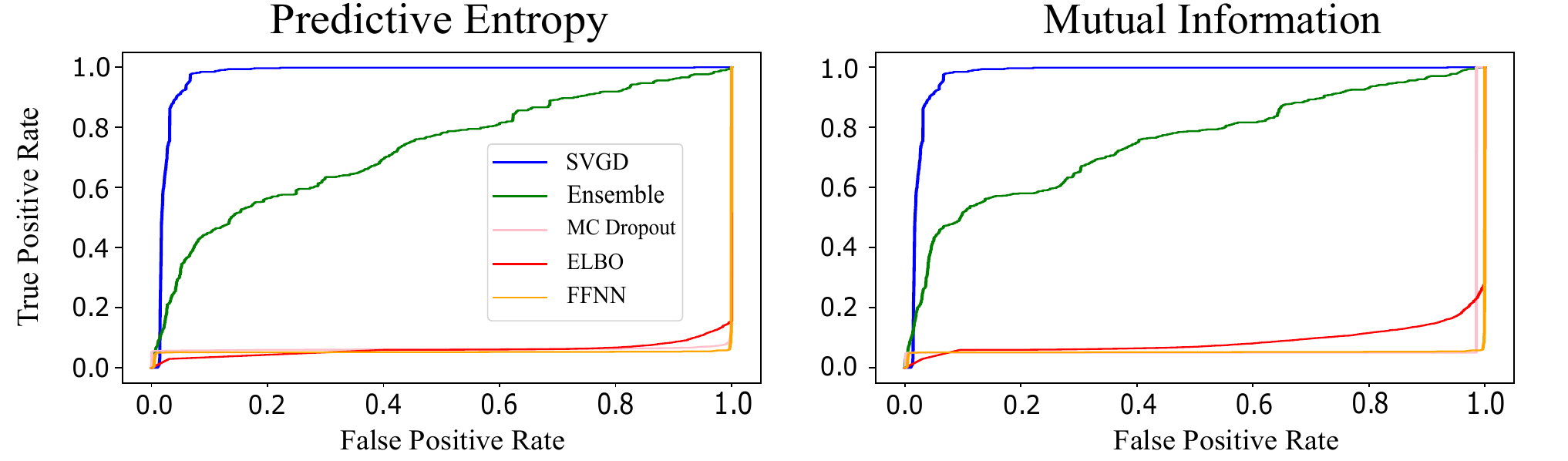}
    \caption{Performance of our proposed method to detect feature space PGD-L1 adversarial Android malware with a budget $\epsilon$ = 60 (FFNN is a \textit{non-Bayesian} baseline).}
    \label{fig:pgd-chart}
    \vspace{-2mm}
\end{figure}

\noindent From Table~\ref{tab:pgd-results}, it shows that feature-space attacks are more potent on the deterministic FFNN network, possibly due to fewer constraints compared to the problem-space \SP attack. For instance, AUC for FFNN dropped from 69.62\% in \SP attacks to 13.56\% in feature-space PGD L1 attacks with $\epsilon$ = 30. Interestingly, Bayesian approaches, except for SVGD, showed decreased effectiveness. We hypothesize that SVGD's repulsive force mechanism fosters diversity and maintains uncertainty, countering strong feature-space attacks like PGD L1. A visualized AUC curve for the attack budget of $\epsilon$ = 60 is shown in Figures \ref{fig:pgd-chart} for Mutual Information and Predictive Entropy, respectively.

\begin{figure}[h]
    \centering
    \includegraphics[width=.9\linewidth]
    {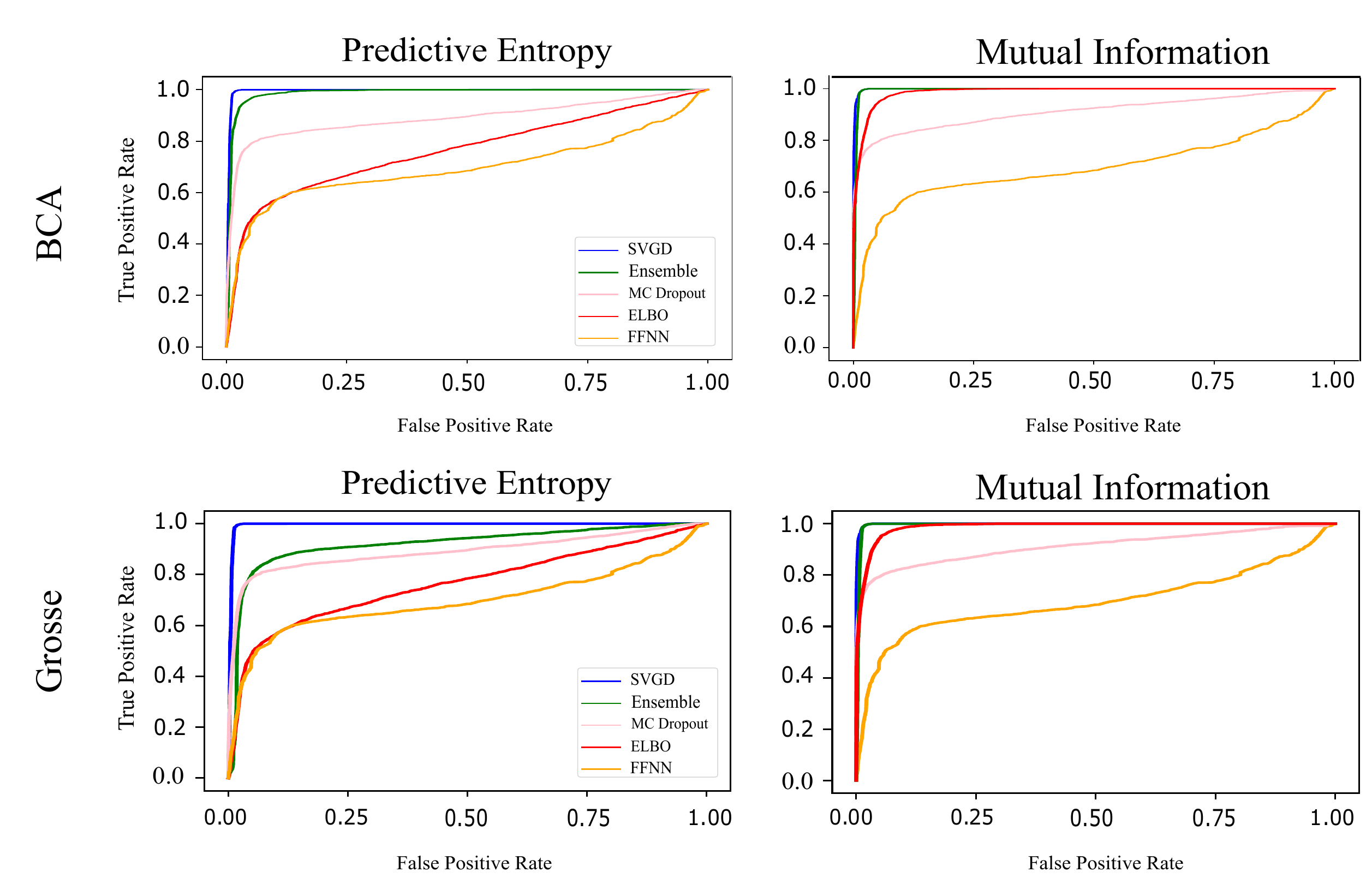}
    \vspace{-1mm}
    \caption{Performance of our proposed method to detect feature space BCA and Grosse adversarial Android malware with a budget $\epsilon$ = 10. (FFNN is a \textit{non-Bayesian} baseline).}
    \label{fig:bca-grosse}
\end{figure}

We also evaluate robustness against feature-space attacks like BCA~\cite{al2018adversarial} and Grosse~\cite{grosse2017adversarial}. In these evaluations, FFNN consistently performs worse than Bayesian approaches, with SVGD demonstrating superior performance, achieving AUC higher than 97\% across all attacking budgets. In addition, our assessment of Grosse attack~\cite{grosse2017adversarial} shows that Bayesian models perform similarly to their counterparts against BCA attacks. Notably, SVGD remains the top-performing model, achieving a minimum AUC of around 96\% across all attack budgets. A visualized AUC curve of both attacks with a budget of $\epsilon$ = 10 is shown in Figures \ref{fig:bca-grosse} for Mutual Information and Predictive Entropy, respectively.

\vspace{-2mm}
\begin{table}[]
\centering
\setlength{\tabcolsep}{5pt}
\resizebox{.95\linewidth}{!}{%
\begin{tabular}{cccccccccccc}
\dtoprule
\multirow{2}{*}{\textbf{\makecell{Networks/\\Attacks}}} & \multirow{2}{*}{$\epsilon$} & \multicolumn{2}{c}{FFNN} & \multicolumn{2}{c}{Dropout} & \multicolumn{2}{c}{ELBO} & \multicolumn{2}{c}{Ensemble} & \multicolumn{2}{c}{SVGD}            \\ 
                                          &                      & PE            & MI       & PE           & MI           & PE          & MI         & PE            & MI           & PE               & MI               \\ \midrule
\multirow{3}{*}{BCA}                      & 5                    & 69.21\%       & NA       & 86.54\%      & 90.32\%      & 70.12\%     & 85.54\%    & 96.12\%       & 97.32\%      & \textbf{97.21\%} & \textbf{98.01\%} \\ 
                                          & 10                   & 72.35\%       & NA       & 89.32\%      & 91.12\%      & 78.43\%     & 96.12\%    & 97.35\%       & 98.94\%      & \textbf{98.45\%} & \textbf{99.12\%} \\ 
                                          & 15                   & 76.45\%       & NA       & 91.15\%      & 92.56\%      & 82.15\%     & 98.43\%    & 98.75\%       & 99.21\%      & \textbf{99.02\%} & \textbf{99.89\%} \\ \midrule
\multirow{3}{*}{Grosse}                   & 5                    & 68.75\%       & NA       & 86.14\%      & 90.22\%      & 69.65\%     & 83.95\%    & 89.12\%       & 96.15\%      & \textbf{96.25\%} & \textbf{97.41\%} \\ 
                                          & 10                   & 72.03\%       & NA       & 88.92\%      & 90.96\%      & 77.42\%     & 95.46\%    & 90.34\%       & 97.95\%      & \textbf{97.85\%} & \textbf{98.05\%} \\ 
                                          & 15                   & 75.95\%       & NA       & 90.54\%      & 91.65\%      & 82.05\%     & 97.68\%    & 92.64\%       & 99.12\%      & \textbf{98.42\%} & \textbf{99.23\%} \\ \dbottomrule
\end{tabular}%
}
\vspace{2mm}
\caption{Detection performance against BCA and Grosse \textit{feature-space} adversarial malware (FFNN is a \textit{non-Bayesian} baseline).}
\label{tab:bca-grosse-results}
\vspace{-15mm}
\end{table}

\subsection{Generalization to PDF malware}
\vspace{-1mm}

Malware detection in PDF files is crucial due to their widespread use. Minor modifications to PDFs, like hidden metadata, can bypass detection systems. PDF malware exploits vulnerabilities, aiming to take control and run malicious code. 
In our experiment, we apply our approach to PDF adversarial malware to test model robustness using the Contagio dataset \cite{Parkour} We employ the unbounded gradient attack method \cite{pdfclassifier} for this experiment.

\begin{figure}[h]
    \centering
    \includegraphics[width=1.0\linewidth]{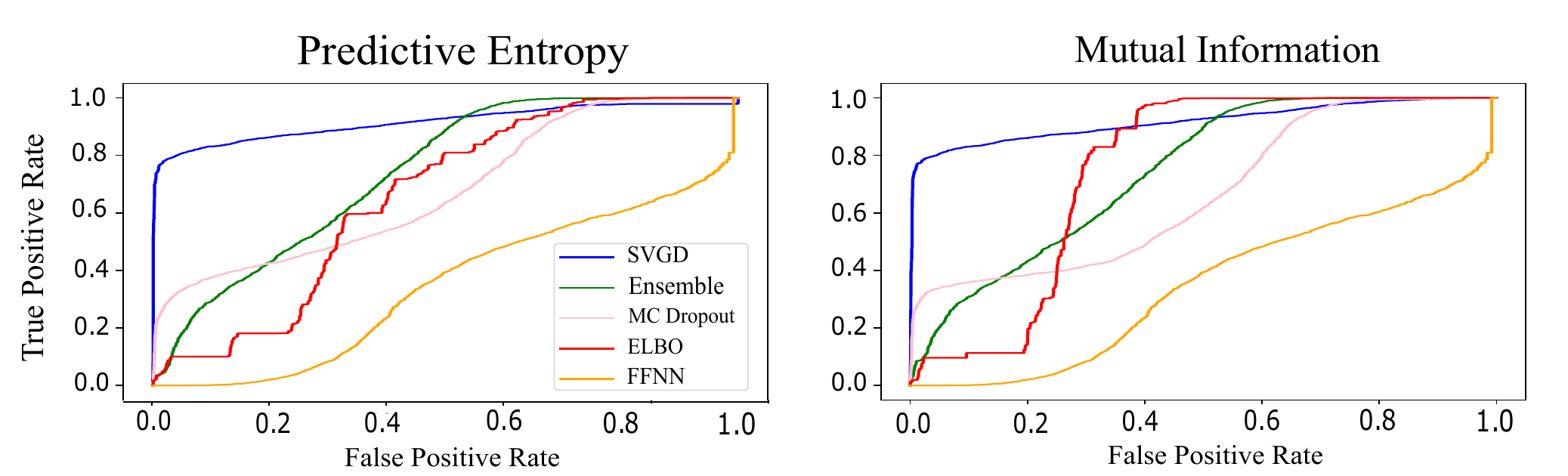}
    \caption{Performance of our proposed method to detect PDF adversarial malware with an attack budget $\epsilon=7$. (FFNN is a \textit{non-Bayesian} baseline).}
    \label{fig:pdf}
\end{figure}

\vspace{2mm}
\noindent\textbf{Results.} Table \ref{tab:pdf-results} shows that the Bayesian approach consistently outperforms single FFNN models, achieving better AUC for both Predictive Entropy and Mutual Information. Notably, SVGD produces the best results among evaluated models, with the highest AUC for both metrics. Therefore, our method can effectively generalize to a different domain such as PDF malware. Figure \ref{fig:pdf} visualizes the AUC curve for the attack budget of $\epsilon=7$ for Mutual Information and Predictive Entropy.

\begin{table}[h]
\centering
\setlength{\tabcolsep}{5pt}
\resizebox{\linewidth}{!}{%
\begin{tabular}{cccccccccccc}
\dtoprule
\multirow{2}{*}{\textbf{\makecell{Networks/\\Attacks}}} & \multirow{2}{*}{$\epsilon$} & \multicolumn{2}{c}{FFNN} & \multicolumn{2}{c}{Dropout} & \multicolumn{2}{c}{ELBO} & \multicolumn{2}{c}{Ensemble} & \multicolumn{2}{c}{SVGD}            \\ 
                                          &                      & PE            & MI       & PE           & MI           & PE          & MI         & PE            & MI           & PE               & MI               \\ \midrule
\multirow{2}{*}{\makecell{Unbounded\\Gradient Attack}}                      & 7                    & 59.43\%       & NA       & 61.12\%      & 60.45\%      & 64.47\%     & 75.12\%    & 71.54\%       & 73.68\%      & \textbf{79.64\%} & \textbf{82.12\%} \\ 
                                          & 8                   & 65.32\%       & NA       & 66.21\%      & 67.46\%      & 69.53\%     & 76.01\%    & 74.75\%       & 76.23\%      & \textbf{91.12\%} & \textbf{92.64\%} \\ \dbottomrule
\end{tabular}%
}
\vspace{2mm}
\caption{Detection performance against PDF adversarial malware (FFNN is a \textit{non-Bayesian} baseline).}
\label{tab:pdf-results}
\vspace{-5mm}
\end{table}

\subsection{Generalization to Windows PE Files}
\noindent This section investigates if our proposed method is able to generalize to an important domain of Windows, namely Windows PE files. 
We focus on Windows PE files because of their popularity and impact. 
We trained FFNN and BNNs with the challenging EMBER~\cite{anderson2018ember} dataset.
Focusing on functional adversarial malware, we use the state-of-the-art problem-space adversarial malware released from~\cite{erdemir2021adversarial}. This released adversarial malware includes 1001 real, functional adversarial malware samples generated using the Greedy Attack method, winner of the DEFCON malware challenge~\cite{Fleshman}. We set adversarial malware as the positive samples and use the benign test set described in Section~\ref{sec:exp-setup} as negative samples. 

\begin{figure}[h]
    \centering
    \includegraphics[width=1.0\linewidth]{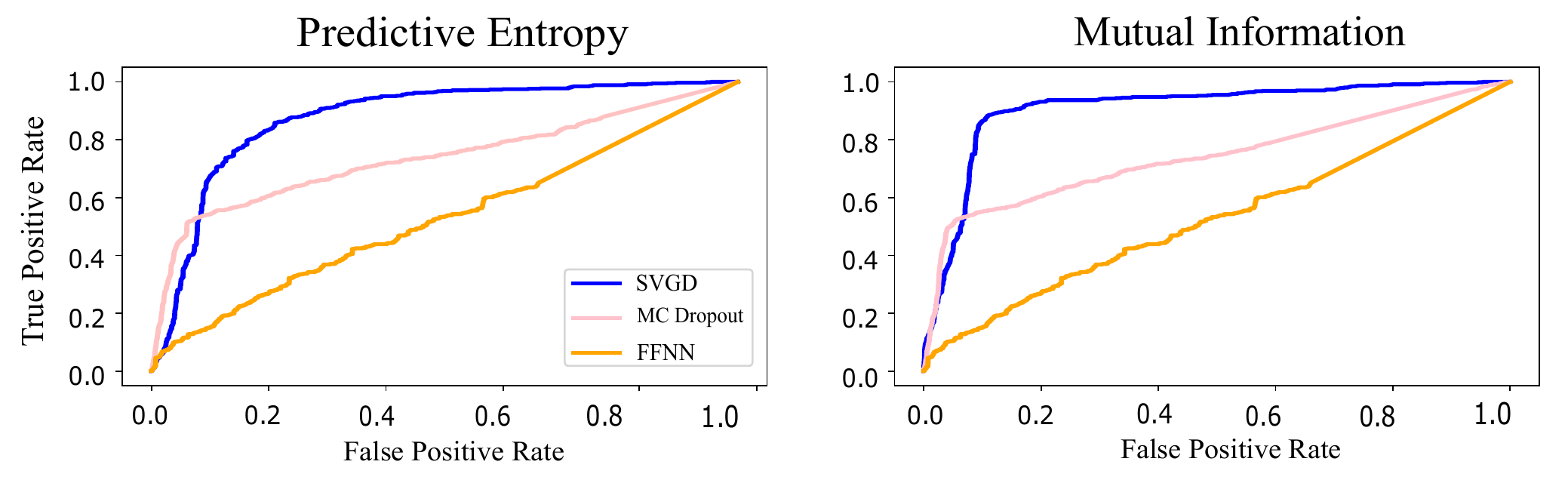}
    \caption{Detection performance against problem-space adversarial Windows PE malware (FFNN is a \textit{non-Bayesian} baseline).}
    \label{fig:windows}
\end{figure}

\vspace{2mm}
\noindent\textbf{Results.~}As shown in Figure~\ref{fig:windows}, the effectiveness of our method. The results are consistent with those in the Android domain, and demonstrate the generalization of our approach across malware domains against realistic adversarial malware.

\section{Identifying Concept Drift}
Data-driven techniques often exhibit bias towards training data, especially pronounced in the malware domain due to \textit{concept drift}~\cite{webb2016characterizing}. Here, malware evolution causes distribution changes over time, posing challenges for ML-based methods affected by the \textit{concept-drift} problem, limiting their applicability.

\begin{figure}[h]
    \centering
    \includegraphics[width=1.0\linewidth]{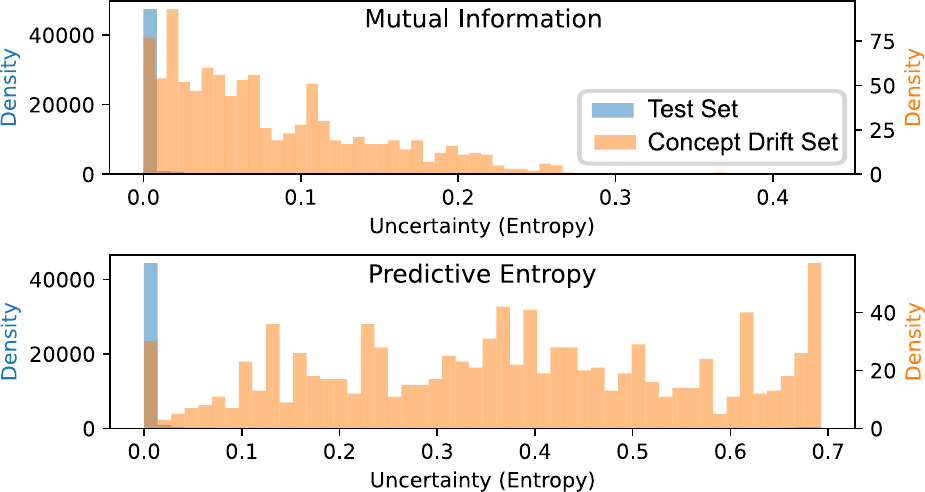}
    \caption{Model diversity-promoting Bayesian methods like SVGD can detect concept drift by measuring uncertainty.}
    \label{fig:drift}
    \vspace{-5mm}
\end{figure}

Our work challenges conventional notions by leveraging uncertainty to detect concept drift, offering a novel perspective. This allows timely detection of evolving malware, prompting prompt retraining or updating of malware detectors. To illustrate, we conducted experiments with Bayesian neural networks trained on the Drebin dataset (Section~\ref{sec:bayes}), containing malware from 2010 to 2012. For concept drift evaluation, we collected a Concept Drift Set with 1K Android malware apps from AndroZoo, spanning 2022 to 2023. Figure~\ref{fig:drift} shows how uncertainty effectively reveals shifts, particularly with the Predictive Entropy measure, aiding in identifying abnormalities for practitioners to notice timely

\vspace{2mm}
\vspace{-3mm}

\section{Model Parameter Diversity Measures}
\label{appd:diversity}

In the absence of a standard measure of the diversity among parameter particles, we propose to use Kullback–Leibler (KL) Divergence between the softmax output of each parameter particle and that of the expected parameters of a Bayesian model to measure the diversity of the models. We compute it over the problem-space adversarial set of $\epsilon=90$ from \SP attack (malware with preserved \textit{realism} and \textit{functionality}). In particular,
$$\text{Diversity}=\frac{1}{N}\sum_{i=1}^N{KL\Big[p(y\mid\bx'_i,\btheta),\mathbb{E}_{\btheta}[p(y\mid\bx'_i,\btheta)]\Big]}$$
where $\text{KL}$ is the Kullback–Leibler divergence, N is the number of samples. 
\begin{figure}[h]
    \centering
    \includegraphics[width=.9\linewidth]{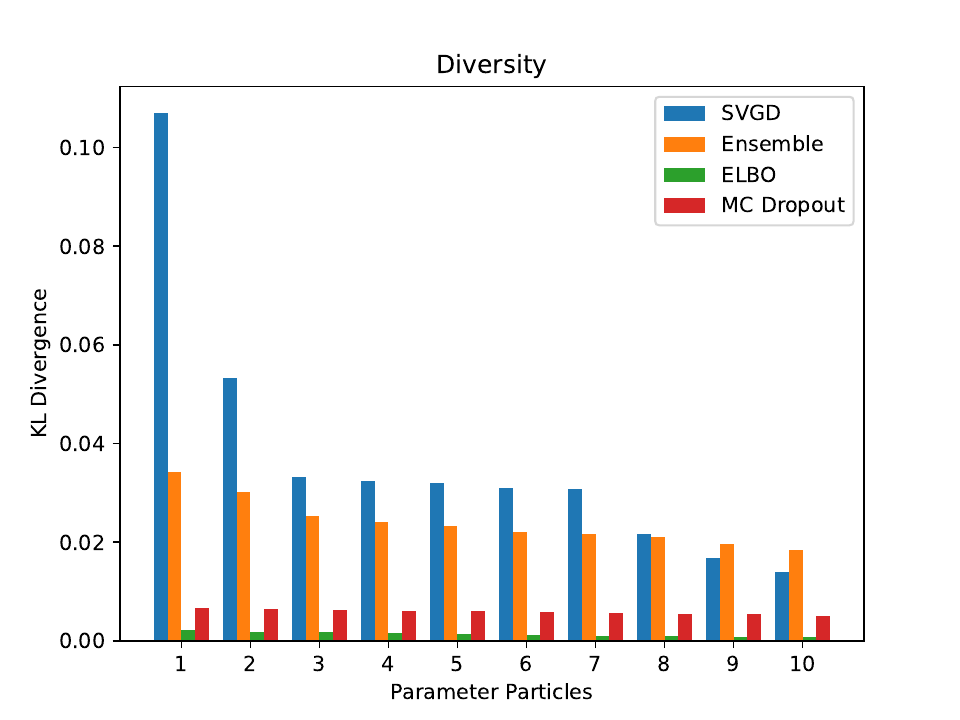}
    \caption{Diversity measures among different learning approaches.}
    \label{fig:diversity}
\end{figure}

\vspace{2mm}
\noindent\textbf{Results.~}Figure~\ref{fig:diversity} shows that the SVGD approach enhances diversity, leading to improved performance in detecting adversarial malware. This supports our notion that diverse models better capture uncertainty, aiding in effective detection. Interestingly, ensemble training, using random initialization seeds, also boosts diversity compared to methods like MC dropout and ELBO. While the ensemble method performs well, it falls short of SVGD's effectiveness, reinforcing the need for improved multi-modal posterior approximation for robust malware defense strategies.

\section{Threat to Validity}
A well-calibrated model is able to assign high probabilities (high confidence or low uncertainty) for benign code and malware but low probabilities (low confidence or high uncertainty) for adversarial malware. In general, evidence show Bayesian neural networks are better calibrated~\cite{NEURIPS2020_b6dfd418} where uncertainty estimates from Bayesian models are consistent with the observed errors. However, due to model under-specifications and approximate inference, uncertainty from Bayesian models can be inaccurate~\cite{NEURIPS2020_b6dfd418,maddox2019simple,foong2019between,yao2019quality}. Interestingly, SVGD approximations in our empirical studies demonstrated the ability to  yield models able to express uncertainty estimates capable of discriminating adversarial malware from benign-ware. Notably, to improve uncertainty estimates, calibration methods can be employed~\cite{detommaso2022uncertainty}. 
\vspace{-2mm}

\section{Conclusion}
\vspace{-1mm}

We propose leveraging efficient and practical approximations of Bayesian neural networks to capture uncertainty better. The approach demonstrated the effectiveness of using uncertainty captured by a probabilistic model to detect adversarial malware without sacrificing performance experienced with adversarial training for robustness (hence, \textit{free}). We have also shown that such techniques allow us to detect concept drift in our data. We do not claim that uncertainty alone provides a strong defense against adversarial malware. However, measuring the uncertainty expressed in the probabilistic model makes it more challenging to attack than its deterministic (single parameter) counterparts. Importantly, the approximation we leverage to learn a BNN, though scalable and more efficient, is still coarse. Our insights suggest that seeking better approximations to capture the posterior is an important avenue for future research to defend against adversarial malware.

\section*{Acknowledgments}

This research was supported by the Next Generation Technologies Fund (NGTF) from the Defence Science and Technology Group (DSTG), Australia. 

\bibliographystyle{splncs04}
\bibliography{main}

\begin{thebibliography}{10}
\providecommand{\url}[1]{\texttt{#1}}
\providecommand{\urlprefix}{URL }
\providecommand{\doi}[1]{https://doi.org/#1}

\bibitem{al2018adversarial}
Al-Dujaili, A., Huang, A., Hemberg, E., O’Reilly, U.M.: Adversarial deep
  learning for robust detection of binary encoded malware. In: IEEE Security
  and Privacy Workshops (S\&PW) (2018)

\bibitem{anderson2018ember}
Anderson, H.S., Roth, P.: {Ember: an open dataset for training static PE
  malware machine learning models}. arXiv preprint arXiv:1804.04637  (2018)

\bibitem{anderson2019measuring}
Anderson, R., Barton, C., B{\"o}hme, R., Clayton, R., Gan{\'a}n, C., Grasso,
  T., Levi, M., Moore, T., Vasek, M.: Measuring the changing cost of
  cybercrime. In: Workshop on the Economics of Information Security (WEIS)
  (2019)

\bibitem{arp2014drebin}
Arp, D., Spreitzenbarth, M., Hubner, M., Gascon, H., Rieck, K., Siemens, C.:
  Drebin: Effective and explainable detection of android malware in your
  pocket. In: Network and Distributed System Security Symposium (NDSS) (2014)

\bibitem{athalye2018obfuscated}
Athalye, A., Carlini, N., Wagner, D.: Obfuscated gradients give a false sense
  of security: Circumventing defenses to adversarial examples. In:
  International Conference on Machine Learning {(ICML)} (2018)

\bibitem{backes17}
Backes, M., Nauman, M.: {{LUNA}}: {{Quantifying}} and {{Leveraging
  Uncertainty}} in {{Android Malware Analysis}} through {{Bayesian Machine
  Learning}}. In: IEEE European Symposium on Security and Privacy (Euro S\&P)
  (2017)

\bibitem{biggio2013evasion}
Biggio, B., Corona, I., Maiorca, D., Nelson, B., {\v{S}}rndi{\'c}, N., Laskov,
  P., Giacinto, G., Roli, F.: Evasion attacks against machine learning at test
  time. In: Joint European Conference on Machine Learning and Knowledge
  Discovery in Databases (ECML PKDD) (2013)

\bibitem{biggio2013security}
Biggio, B., Fumera, G., Roli, F.: Security evaluation of pattern classifiers
  under attack. IEEE Transactions on Knowledge and Data Engineering
  \textbf{26}(4),  984--996 (2013)

\bibitem{biggio2018wild}
Biggio, B., Roli, F.: Wild patterns: Ten years after the rise of adversarial
  machine learning. Pattern Recognition  \textbf{84},  317--331 (2018)

\bibitem{blundell2015weight}
Blundell, C., Cornebise, J., Kavukcuoglu, K., Wierstra, D.: {Weight uncertainty
  in neural network}. In: {International Conference on Machine Learning (ICML)}
  (2015)

\bibitem{carlini2017towards}
Carlini, N., Wagner, D.: Towards evaluating the robustness of neural networks.
  In: IEEE Symposium on Security and Privacy (S\&P) (2017)

\bibitem{chen2019android}
Chen, X., Li, C., Wang, D., Wen, S., Zhang, J., Nepal, S., Xiang, Y., Ren, K.:
  Android hiv: A study of repackaging malware for evading machine-learning
  detection. IEEE Transactions on Information Forensics and Security (TIFS)
  \textbf{15},  987--1001 (2019)

\bibitem{pdfclassifier}
Chen, Y., Wang, S., She, D., Jana, S.: On training robust {PDF} malware
  classifiers. In: USENIX Conference on Security Symposium (2020)

\bibitem{croce2022sparse}
Croce, F., Andriushchenko, M., Singh, N.D., Flammarion, N., Hein, M.:
  Sparse-rs: a versatile framework for query-efficient sparse black-box
  adversarial attacks. In: AAAI Conference on Artificial Intelligence (2022)

\bibitem{d2021stein}
D'Angelo, F., Fortuin, V., Wenzel, F.: On stein variational neural network
  ensembles. In: International Conference on Machine Learning (ICML) Workshop
  on Uncertainty and Robustness in Deep Learning (2021)

\bibitem{DEFCON}
DEFCON: Machine learning static evasion competition.
  \url{https://www.elastic.co/blog/machine-learning-static-evasion-competition}
  (2019), accessed: 2022-08-09

\bibitem{demetrio2021secml}
Demetrio, L., Biggio, B.: Secml-malware: Pentesting windows malware classifiers
  with adversarial exemples in python. arXiv preprint arXiv:2104.12848  (2021)

\bibitem{demontis2017yes}
Demontis, A., Melis, M., Biggio, B., Maiorca, D., Arp, D., Rieck, K., Corona,
  I., Giacinto, G., Roli, F.: Yes, machine learning can be more secure! a case
  study on android malware detection. IEEE Transactions on Dependable and
  Secure Computing (TDSC)  \textbf{16}(4),  711--724 (2019)

\bibitem{detommaso2022uncertainty}
Detommaso, G., Gasparin, A., Wilson, A., Archambeau, C.: Uncertainty
  calibration in bayesian neural networks via distance-aware priors. arXiv
  preprint arXiv:2207.08200  (2022)

\bibitem{doan2022bayesian}
Doan, B.G., Abbasnejad, E.M., Shi, J.Q., Ranasinghe, D.C.: Bayesian learning
  with information gain provably bounds risk for a robust adversarial defense.
  In: International Conference on Machine Learning (ICML) (2022)

\bibitem{doan2023featurespace}
Doan, B.G., Yang, S., Montague, P., Vel, O.D., Abraham, T., Camtepe, S.,
  Kanhere, S.S., Abbasnejad, E., Ranasinghe, D.C.: Feature-space bayesian
  adversarial learning improved malware detector robustness. In: AAAI
  Conference on Artificial Intelligence (2023)

\bibitem{eddy_perlroth_2020}
Eddy, M., Perlroth, N.:
  \url{https://www.nytimes.com/2020/09/18/world/europe/cyber-attack-germany-ransomeware-death.html}
  (Sep 2020), accessed: 2022-12-01

\bibitem{erdemir2021adversarial}
Erdemir, E., Bickford, J., Melis, L., Aydore, S.: Adversarial robustness with
  non-uniform perturbations. In: Advances in Neural Information Processing
  Systems (NeurIPS) (2021)

\bibitem{feinman2017detecting}
Feinman, R., Curtin, R.R., Shintre, S., Gardner, A.B.: Detecting adversarial
  samples from artifacts. arXiv preprint arXiv:1703.00410  (2017)

\bibitem{Fleshman}
Fleshman, W.: Evading machine learning malware classifiers.
  \url{https://towardsdatascience.com/evading-machine-learning-malware-classifiers-ce52dabdb713}
  (2019), accessed: 2022-08-09

\bibitem{NEURIPS2020_b6dfd418}
Foong, A., Burt, D., Li, Y., Turner, R.: On the expressiveness of approximate
  inference in bayesian neural networks. In: Advances in Neural Information
  Processing Systems (NeurIPS). pp. 15897--15908 (2020)

\bibitem{foong2019between}
Foong, A.Y., Li, Y., Hern{\'a}ndez-Lobato, J.M., Turner, R.E.: {`In-Between'
  Uncertainty in Bayesian Neural Networks}. arXiv preprint arXiv:1906.11537
  (2019)

\bibitem{gal2016uncertainty}
Gal, Y., et~al.: Uncertainty in deep learning  (2016)

\bibitem{goodfellowExplainingHarnessingAdversarial2015}
Goodfellow, I.J., Shlens, J., Szegedy, C.: Explaining and harnessing
  adversarial examples. In: {{International Conference}} on {{Learning
  Representations}} {{(ICLR)}} (2015)

\bibitem{grosse2016adversarial}
Grosse, K., Papernot, N., Manoharan, P., Backes, M., McDaniel, P.: Adversarial
  perturbations against deep neural networks for malware classification. arXiv
  preprint arXiv:1606.04435  (2016)

\bibitem{grosse2017adversarial}
Grosse, K., Papernot, N., Manoharan, P., Backes, M., McDaniel, P.: Adversarial
  examples for malware detection. In: European Symposium on Research in
  Computer Security (ESORICS) (2017)

\bibitem{harang2020sorel}
Harang, R., Rudd, E.M.: {SOREL-20M}: A large scale benchmark dataset for
  malicious pe detection (2021)

\bibitem{kaspersky}
KasperskyLab: Cybercriminals attack users with 411,000 new malicious files
  daily.
  \url{https://www.kaspersky.com/about/press-releases/2023_rising-threats-cybercriminals-unleash-411000-malicious-files-daily-in-2023}
  (2023), accessed: 2024-01-09

\bibitem{kolosnjaji2018adversarial}
Kolosnjaji, B., Demontis, A., Biggio, B., Maiorca, D., Giacinto, G., Eckert,
  C., Roli, F.: Adversarial malware binaries: Evading deep learning for malware
  detection in executables. In: European Signal Processing Conference (EUSIPCO)
  (2018)

\bibitem{kreuk2018deceiving}
Kreuk, F., Barak, A., Aviv-Reuven, S., Baruch, M., Pinkas, B., Keshet, J.:
  Deceiving end-to-end deep learning malware detectors using adversarial
  examples. arXiv preprint arXiv:1802.04528  (2018)

\bibitem{krishnan2022bayesiantorch}
Krishnan, R., Esposito, P., Subedar, M.: Bayesian-torch: Bayesian neural
  network layers for uncertainty estimation.
  \url{https://github.com/IntelLabs/bayesian-torch} (2022)

\bibitem{li2020adversarial}
Li, D., Li, Q.: Adversarial deep ensemble: Evasion attacks and defenses for
  malware detection. IEEE Transactions on Information Forensics and Security
  (TIFS)  \textbf{15},  3886--3900 (2020)

\bibitem{li2021framework}
Li, D., Li, Q., Ye, Y., Xu, S.: A framework for enhancing deep neural networks
  against adversarial malware. IEEE Transactions on Network Science and
  Engineering  \textbf{8}(1),  736--750 (2021)

\bibitem{li2021can}
Li, D., Qiu, T., Chen, S., Li, Q., Xu, S.: Can we leverage predictive
  uncertainty to detect dataset shift and adversarial examples in android
  malware detection? In: Annual Computer Security Applications
  Conference(ACSAC) (2021)

\bibitem{li2017dropout}
Li, Y., Gal, Y.: Dropout inference in bayesian neural networks with
  alpha-divergences. In: International Conference on Machine Learning (ICML)
  (2017)

\bibitem{Liu2016}
Liu, Q., Wang, D.: Stein variational gradient descent: A general purpose
  bayesian inference algorithm. In: Advances in Neural Information Processing
  Systems (NeurIPS) (2016)

\bibitem{mackay1992practical}
MacKay, D.J.C.: {A Practical Bayesian Framework for Backpropagation Networks}.
  Neural Computation  \textbf{4}(3),  448--472 (05 1992)

\bibitem{maddox2019simple}
Maddox, W.J., Izmailov, P., Garipov, T., Vetrov, D.P., Wilson, A.G.: A simple
  baseline for bayesian uncertainty in deep learning. In: Advances in Neural
  Information Processing Systems (NeurIPS) (2019)

\bibitem{pgd}
Madry, A., Makelov, A., Schmidt, L., Tsipras, D., Vladu, A.: Towards deep
  learning models resistant to adversarial attacks. In: International
  Conference on Learning Representations (ICLR) (2018)

\bibitem{melis2019secml}
Melis, M., Demontis, A., Pintor, M., Sotgiu, A., Biggio, B.: Secml: A python
  library for secure and explainable machine learning. arXiv preprint
  arXiv:1912.10013  (2019)

\bibitem{nguyenLeveragingUncertaintyImproved2021}
Nguyen, A.T., Raff, E., Nicholas, C., Holt, J.: Leveraging {{Uncertainty}} for
  {{Improved Static Malware Detection Under Extreme False Positive
  Constraints}}. In: {International Joint Conferences on Artificial
  Intelligence (IJCAI) Workshop} (2021)

\bibitem{Parkour}
Parkour, M.: 16,800 clean and 11,960 malicious files for signature testing and
  research.,
  \url{https://contagiodump.blogspot.com/2013/03/16800-clean-and-11960-malicious-files.html}

\bibitem{paszke2019pytorch}
Paszke, A., Gross, S., Massa, F., Lerer, A., Bradbury, J., Chanan, G., Killeen,
  T., Lin, Z., Gimelshein, N., Antiga, L., et~al.: Pytorch: An imperative
  style, high-performance deep learning library (2019)

\bibitem{peng2012using}
Peng, H., Gates, C., Sarma, B., Li, N., Qi, Y., Potharaju, R., Nita-Rotaru, C.,
  Molloy, I.: Using probabilistic generative models for ranking risks of
  android apps. In: ACM Conference on Computer and Communications Security
  (CCS) (2012)

\bibitem{pierazzi2020intriguing}
Pierazzi, F., Pendlebury, F., Cortellazzi, J., Cavallaro, L.: Intriguing
  properties of adversarial ml attacks in the problem space. In: IEEE Symposium
  on Security and Privacy (S\&P) (2020)

\bibitem{raff2018malware}
Raff, E., Barker, J., Sylvester, J., Brandon, R., Catanzaro, B., Nicholas,
  C.K.: Malware detection by eating a whole exe. In: AAAI Conference on
  Artificial Intelligence Workshop (2018)

\bibitem{rawat2017adversarial}
Rawat, A., Wistuba, M., Nicolae, M.I.: Adversarial phenomenon in the eyes of
  bayesian deep learning. arXiv preprint arXiv:1711.08244  (2017)

\bibitem{ritter2018scalable}
Ritter, H., Botev, A., Barber, D.: A scalable laplace approximation for neural
  networks. In: International Conference on Learning Representations (ICLR)
  (2018)

\bibitem{smith2018understanding}
Smith, L., Gal, Y.: Understanding measures of uncertainty for adversarial
  example detection. In: Uncertainty in Artificial Intelligence (UAI) (2018)

\bibitem{srivastava2014dropout}
Srivastava, N., Hinton, G., Krizhevsky, A., Sutskever, I., Salakhutdinov, R.:
  Dropout: a simple way to prevent neural networks from overfitting. Journal of
  Machine Learning Research (JMLR)  \textbf{15}(1),  1929--1958 (2014)

\bibitem{suciu2019exploring}
Suciu, O., Coull, S.E., Johns, J.: Exploring adversarial examples in malware
  detection. In: IEEE Security and Privacy Workshops (S\&PW) (2019)

\bibitem{webb2016characterizing}
Webb, G., Hyde, R., Cao, H., Nguyen, H.L., Petitjean, F.: Characterizing
  concept drift. Data Mining and Knowledge Discovery  \textbf{30},  964--994
  (2016)

\bibitem{yang2017malware}
Yang, W., Kong, D., Xie, T., Gunter, C.A.: Malware detection in adversarial
  settings: Exploiting feature evolutions and confusions in android apps. In:
  Annual Computer Security Applications Conference (ACSAC) (2017)

\bibitem{yao2019quality}
Yao, J., Pan, W., Ghosh, S., Doshi-Velez, F.: Quality of uncertainty
  quantification for bayesian neural network inference. In: International
  Conference on Machine Learning (ICML) Workshop on Uncertainty \& Robustness
  in Deep Learning (2019)

\end{thebibliography}

\end{document}